\documentstyle[aps,prl,multicol,psfig,amsmath]{revtex}
\begin{document}

\input{epsf}

\title{Superfluidity in a Doped Helium Droplet}
\author{E. W. Draeger$^\dag$ and  D. M. Ceperley}
\address{Department of Physics and National Center for Supercomputing
Applications, University of Illinois Urbana-Champaign, 61801}
\maketitle

\begin{abstract}
Path Integral Monte Carlo calculations of the superfluid density
throughout $^4$He droplets doped with linear impurities (HCN)$_n$
are presented.  After deriving a local estimator for the superfluid
density distribution, we find a decreased superfluid response in
the first solvation layer.  This effective normal fluid exhibits
temperature dependence similar to that of a two-dimensional helium
system.
\end{abstract}

\begin{multicols}{2}
\narrowtext

Molecules confined in helium nanodroplets have been shown to exhibit
excitation spectra with clearly resolved rotational fine structure
consistent with that of a free rotor, though with an increased moment
of inertia.  Grebenev \emph{et al.}\cite{grebenev98} have shown that
only a few layers of $^4$He coating the molecule are required to
decouple the impurity from the droplet and achieve free rotation.
Callegari \emph{et al.}\cite{callegari99} have suggested that the
increased moment of inertia is due to the hydrodynamic response of the
impurity rotating through the anisotropic helium density immediately
surrounding the molecule.  Kwon and Whaley\cite{kwon99} have proposed
a model in which a microscopic normal fluid is induced in the first
solvation layer by the anisotropy of the molecule-helium interaction.

In this paper, a microscopic Path Integral Monte Carlo (PIMC)
estimator of the local contribution to the total superfluid
response is presented.  Simulations of $^4$He droplets containing
linear (HCN)$_3$ isomers show a significant reduction in the
superfluid response of the first solvation layer.  Furthermore, we
find that this reduction is due in part to thermal excitations, in
addition to the temperature-independent normal fluid assumed to be
induced by the anisotropic HCN-helium interaction potential.  The
cylindrically-symmetric region of the first solvation layer was
found to exhibit temperature dependence consistent with that of a
two-dimensional helium system.  The moment of inertia was
calculated, in reasonable agreement with experiment, and was
independent of temperature below T=1~K as the dominant
contribution came from the normal fluid coating the ends of the
molecule, which was predominantly induced by the anisotropy of the
density in that region.  

The superfluid density can be defined in terms of the response of
the system to an imposed rotation.  In imaginary-time path
integrals, it is manifested by particle exchange over length
scales equal to the system size.  Although superfluid response
(like conductivity) is not itself a local property, it is possible
to calculate a local contribution to the total response.
In PIMC, the total superfluid response along the axis of rotation
$\hat{n}$ is proportional to the square of total projected area of
the imaginary-time paths\cite{pollock87}:
\begin{equation}
\left. \frac{\rho_s}{\rho} \right|_{\hat{n}} = \frac{2 m
\left\langle A_{\hat{n}}^2 \right\rangle }{\beta \lambda I_c},
\end{equation}
where $\lambda = \hbar^2/2m$, $\beta = 1/k_BT$ and $I_c$ is the
classical moment of inertia. To define a local superfluid density
we write:
\begin{align}
\left. \rho_s({\bf r})\right|_{\hat{n}} = & \frac{2 m
N}{\beta\lambda I_c} \left\langle \int d{\bf r'} \,
A_{\hat{n}}({\bf r}) A_{\hat{n}}({\bf r'})
\right\rangle \nonumber \\
= & \frac{2 m N}{\beta\lambda I_c} \left\langle A_{\hat{n}}({\bf
r}) A_{\hat{n}} \right\rangle, \label{pimclocalsf1}
\end{align}
where ${\bf A}({\bf r})$, related to the local angular momentum
operator, is defined as
\begin{equation}
{\bf A}({\bf r}) \equiv \frac{1}{2}\sum\limits_{i,k} ({\bf r}_{i,k}
\times {\bf r}_{i,k+1}) \; \delta({\bf r}-{\bf r}_{i,k})
\label{pimclocalsf2}
\end{equation}
and $A_{\hat{n}}$ is the ${\hat{n}}$-component of the total area
of all the particles. Since ${\bf A}({\bf r})$ integrates to the
total area, the local superfluid response exactly integrates to
the total response.

\begin{figure}[hbt]
\epsfxsize=\columnwidth
\centerline{\epsfbox{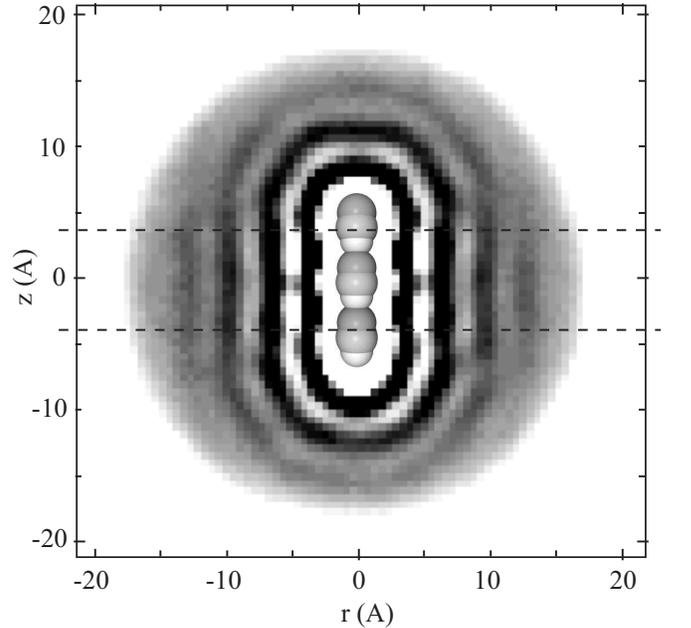}}
\caption{Average number density of an $N=500$ $^4$He-(HCN)$_3$ droplet
at $T=0.38$~K.  The grey scale saturates at $\rho=0.03$~\AA$^{-3}$
(black).  The dashed lines define the cylindrically-symmetric
region used for averaging superfluid density.}
\label{hcnden}
\end{figure}

Two types of contributions to the local superfluid density can be
distinguished based on the connectivity of the instantaneous
paths: contributions of particles on the same chain, which on the
average must be positive, and contributions of particles on
different exchange cycles.  By reversing the order of one exchange
cycle the contribution from different cycles will change sign; if
the cycles are spatially separated, the magnitude of the
contribution will be unaffected, so that their contribution is
much smaller. They will however increase the statistical noise of
the superfluid density. Numerically we find that the same cycle
contribution accounts for roughly 80\% of the total superfluid
density in the systems presented in this paper.

The local superfluid density estimator used by Kwon and
Whaley\cite{kwon99} defined the effective normal fluid induced by
the anisotropic molecule-helium interaction potential in terms of
the average number density distribution of paths in exchange
cycles of fewer than six atoms. Although qualitatively
interesting, this estimator is not the superfluid response to an
imposed rotation, as there is no direct relation between the
number of atoms in a permutation cycle and its area.

In order to test our estimator, PIMC simulations were performed on
$N=128$ and $N=500$ $^4$He droplets doped with (HCN)$_3$ isomers.
Nauta and Miller\cite{nauta99} found that HCN molecules in helium
droplets self-assemble into linear chains spaced roughly 4.4~\AA\
apart. The HCN molecules in our simulations were fixed along the
z-axis with this spacing.  An imaginary time-step of
$1/20$~K$^{-1}$ was used.  With cylindrical symmetry and precise
experimental data over a range of isomer lengths, this system is
well-suited for studying superfluidity at a molecular interface.

\begin{figure}[hbt]
\epsfxsize=\columnwidth
\centerline{\epsfbox{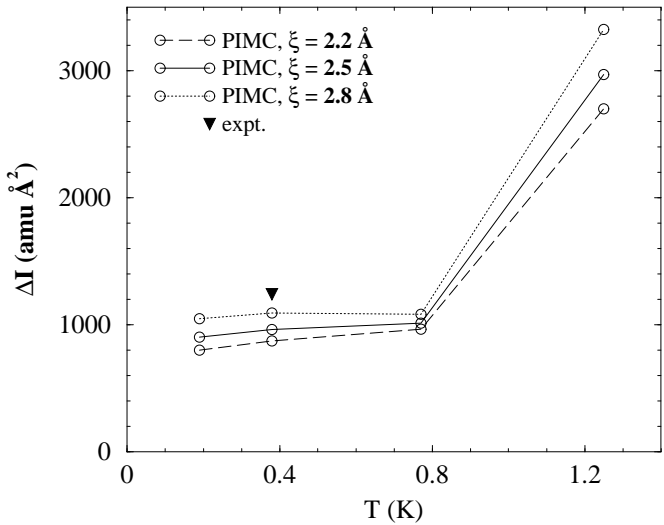}}
\vskip0.50cm \caption {Temperature dependence of moment of inertia
difference of (HCN)$_3$ isomer from the gas phase value due to the
helium droplet.  Values were calculated from Eq.~(\ref{moi_pimc1}),
using the local superfluid density distributions calculated from
$N=128$ PIMC simulations.  Also shown is the experimental value for
(HCN)$_3$.}
\label{moi_vd90a_twofluid}
\end{figure}

The number density of a doped helium droplet at $T=0.38$~K is
shown in Fig.~\ref{hcnden}.  The two-dimensional areal density of
the first solvation layer was $0.12$~\AA$^{-2}$, which is still in
the liquid phase for thin $^4$He films at these temperatures
\cite{bishop81}. Even if the density were large enough to be solid
in a 2D film,  the curved geometry of the film around the HCN
molecule could frustrate crystalline order, keeping the first
solvation layer in a dense liquid state.

\begin{figure}[hbt]
\epsfxsize=\columnwidth
\centerline{\epsfbox{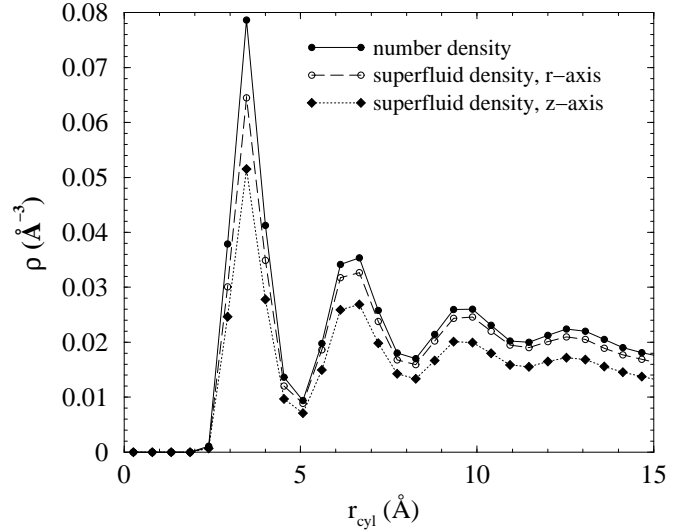}}
\caption{Superfluid and number density distributions, averaged over
the cylindrically-symmetric region of an $N=500$ $^4$He-(HCN)$_3$
droplet at $T=0.38$~K.}
\label{avgsf_vd41}
\end{figure}

We define the change in the moment of inertia due to the helium as
the contribution due to the normal fraction in the first layer
rotating rigidly with the impurity:
\begin{equation}
\Delta I = \int_{v(\xi)} d{\bf r} \, m_4 \, r_\perp^2 \, \left(
\rho({\bf r}) - \rho_{s}({\bf r})\right) \label{moi_pimc1}
\end{equation}
where $v(\xi)$ is the volume of helium a distance $\xi$ away from the
surface of the molecule, $r_\perp$ is the radial distance from the
axis of rotation in cylindrical coordinates.  The results calculated
from our $N=128$ droplet results are plotted in
Fig.~\ref{moi_vd90a_twofluid}, for several cutoff distances near the
first layer minimum.  Both the estimated statistical error and the
uncertainty due to the cut-off were on the order of 10\%.
Fig.~\ref{moi_vd90a_twofluid} shows that the moment of inertia due to
the effective normal fluid in the first solvation layer calculated
using our PIMC local superfluid density estimator is in reasonable
agreement with the experimental value of $\Delta I =
1240$~amu~\AA$^{2}$ \cite{nauta01}, within error bars.  We also find
that $\Delta I$ is effectively independent of temperature below
%
%
$T=1.0$~K.  This is because the dominant contribution to the moment of inertia
comes from the induced normal fluid at the ends of the isomer, which
is almost completely due to anisotropy in the molecular potential.

\begin{figure}[hbt]
\epsfxsize=\columnwidth
\centerline{\epsfbox{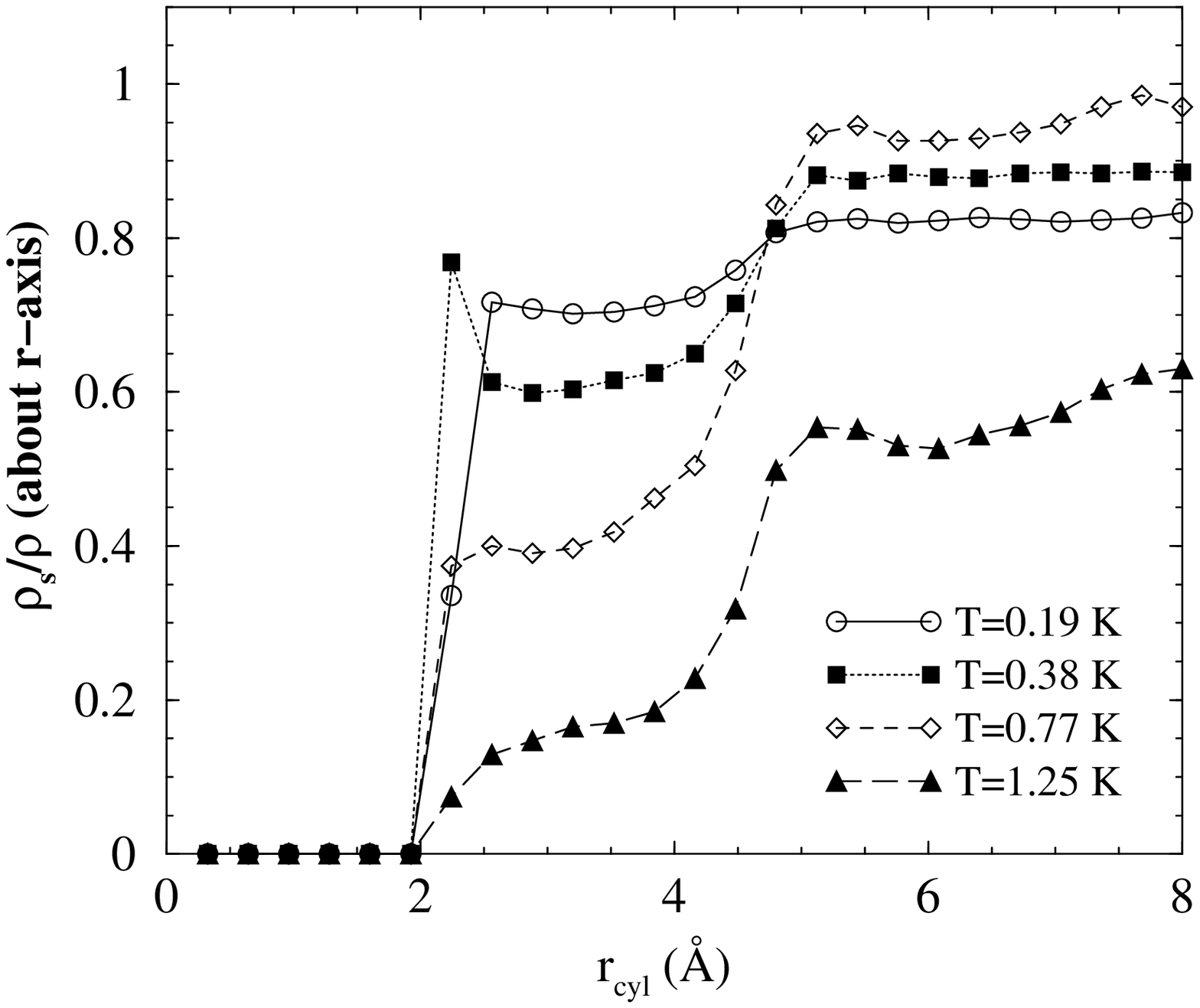}}
\epsfxsize=\columnwidth
\centerline{\epsfbox{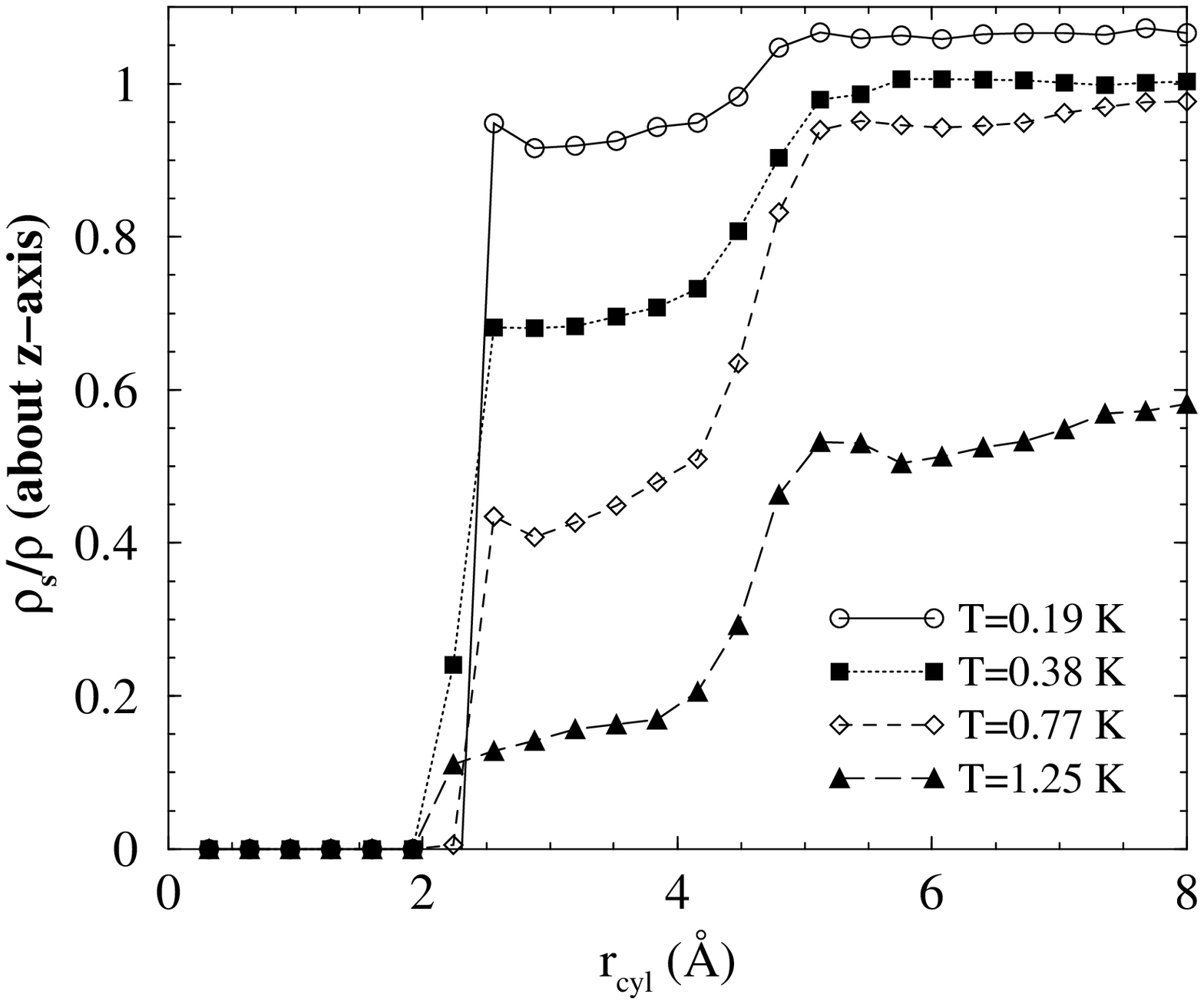}} \vskip0.50cm
\caption {Superfluid density fraction distributions, averaged over the
cylindrically-symmetric region of an $N=128$ $^4$He-(HCN)$_3$ droplet
at $T=0.19$, 0.38, 0.77, and 1.25~K.  The top graph shows superfluid
response about the radial axis, the bottom graph shows superfluid
response about the molecule axis.}
\label{avgratio_vd90a}
\end{figure}

To quantify the reduced superfluid response in the first layer, we
averaged over the cylindrically-symmetric region of the
$^4$He-(HCN)$_3$ droplet, defined as the region between
$z=-3.5$~\AA\ and $z=3.5$~\AA\ (see Fig.~\ref{hcnden}).  The
results of this averaging are shown in Fig.~\ref{avgsf_vd41}.
Taking the ratio of the superfluid density to the number density
clearly shows a reduced superfluid response in the first layer for
rotation about both the radial axis and the molecular axis.  At
zero temperature, there can be no ``normal'' density for rotation
about the molecular axis because of the cylindrical symmetry, so this
reduction must be due to thermal excitations.

Shown in Fig.~\ref{avgratio_vd90a} is the temperature dependence of
the superfluid density in the first solvation layer. determined for
$N=128$ $^4$He-(HCN)$_3$ droplets at $T=0.19$, 0.38, 0.77, and
1.25~K. At first glance, this result appears to contradict known
properties of liquid helium; the superfluid density of bulk helium at
0.75~K is 1.000(1). Theoretical studies of pure $N=128$ helium
clusters show a superfluid transition roughly in agreement with the
bulk lambda transition\cite{sindzingre89}. Pure droplets like those
produced for use in scattering experiments, with several thousand atoms
at $T=0.38$~K, should have a superfluid fraction very close to unity.
However, the helium in the first layer coating the impurity molecule
more closely resembles a two-dimensional system than a
three-dimensional system, because the motion of the helium atoms is
restricted by the He-HCN potential to be primarily on the cylindrical
surface circumscribing the (HCN)$_3$ impurity.  Two-dimensional helium
films have been shown to have a Kosterlitz-Thouless type of superfluid
transition at temperatures significantly lower than $T_\lambda$
\cite{2dsd,nyeki95}.  The reduction in the transition temperature is
due to the reduced dimensionality, increasing the phase space for long
wavelength fluctuations, and the lowering of the ``roton'' gap.  A
%
%
similar temperature-dependent reduction in the superfluid response of
the first layer of helium surrounding the ends of the isomer was not
observed.

\begin{figure}[hbt]
\epsfxsize=\columnwidth
\centerline{\epsfbox{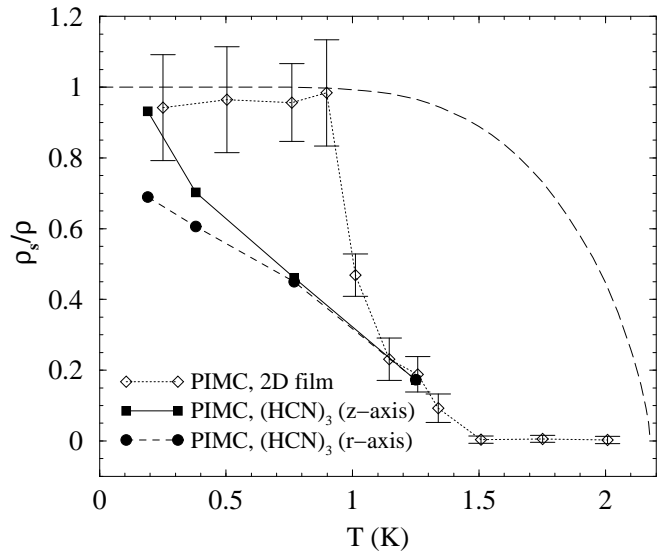}} \vskip0.50cm
\caption {Superfluid density fraction of the cylindrically-symmetric
portion of the first layer of an $N=128$ $^4$He-(HCN)$_3$ droplet
vs. temperature, calculated using Eq.~(\ref{sflayer1}).  The first
layer in these systems has an estimated coverage of
$\sigma=0.12$~\AA$^{-2}$.  Superfluid response about both the molecule
axis (solid squares) and radial axis (solid circles) are presented.
Also shown are the PIMC results of Gordillo et al (open diamonds), for
a 2D $^4$He film with $\sigma=0.051$~\AA$^{-2}$.  For reference, the
superfluid transition in bulk $^4$He is included (dashed line).}
\label{sfvstemp_vd90a}
\end{figure}

To extract the average superfluid density fraction in the first layer
from the distributions plotted in Fig.~\ref{avgratio_vd90a}, we
integrate over the superfluid density and number density in the first
layer and take the ratio:
\begin{equation}
\frac{\rho_s}{\rho} = \frac{\int\limits_0^{r_1} dr_{cyl} \, \rho_s(r_{cyl})}{\int\limits_0^{r_1} dr_{cyl} \, \rho(r_{cyl})},
\label{sflayer1}
\end{equation}
where $r_1$ is the position of the density minimum between the first
and second solvation layers.  The average superfluid density in the
first solvation layer as a function of temperature is shown in
Fig.~\ref{sfvstemp_vd90a}.  The transition is very broad due to the
small number of atoms ($\sim 30$) in the first layer.  It shows the
onset of superfluidity at roughly 1~K. The superfluid density about
the molecule axis (z-axis) goes to unity as the temperature goes to
zero, as the density is symmetric about this axis.  The normal fluid
response from rotating about the radial axis (r-axis) combines both
thermal effects and effective normal fluid induced by rotating through
the anisotropic helium density, and because of this, does not go to
zero at $T=0$~K\cite{callegari99}.

\begin{figure}[hbt]
\epsfxsize=\columnwidth
\centerline{\epsfbox{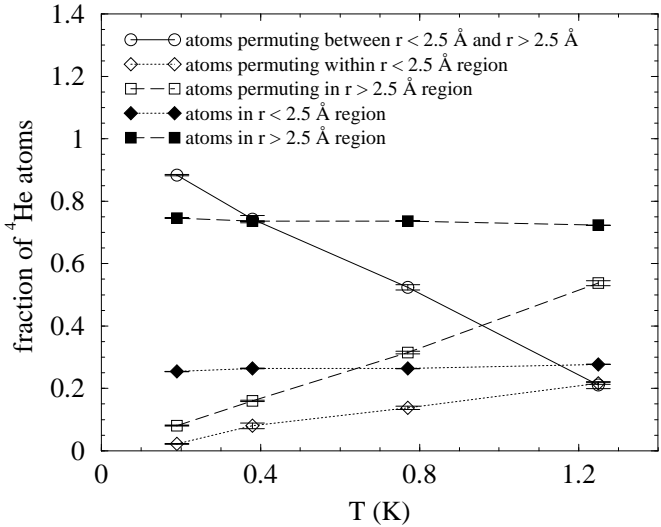}} \vskip0.50cm
\caption {Exchange behavior of the first layer (defined as 2.5~\AA\ or less
from the surface of the (HCN)$_3$ isomer) as a function of
temperature.  A permutation cycle was considered to be part of the
first layer if more than $M/2$ consecutive time slices were
contained within it, where $M$ is the number of imaginary time slices
per atom.}
\label{exchange_comp}
\end{figure}

We have determined that only a small fraction of the particles in the
first layer are localized (not permuting) at T=0.38~K and below, as
shown in Fig.~\ref{exchange_comp}.  Though many of the permutations
are between atoms within the same layer, the first layer is not cut
off from the rest of the fluid.  Below 1~K, the majority of the atoms
in the droplet are part of exchange cycles with atoms in both the
first layer and the rest of the droplet.  However, in terms of
excitations, the first layer is well represented as a 2D superfluid.

To compare the effects of density on the thermally-induced normal
fluid in the first layer, we calculated the superfluid density
distribution for $N=128$ $^4$He-(HCN)$_3$ droplets at T=0.19, 0.38,
0.77, and 1.25~K, with a $^4$He-HCN interaction reduced by a factor of
two in order to reduce the density in the first solvation layer.  The
density in the first layer decreased by $\sim 15$\%, corresponding to
an average coverage of 0.10~\AA$^{-2}$.  The difference in coverage
caused a 20\% reduction in the superfluid response about the radial
axis at T=0.19~K and T=0.38~K.  At higher temperatures, the superfluid
response was unchanged within the estimated error bars.  The
superfluid response about the molecule axis (z-axis) was unchanged
within error bars.  This is further evidence that the normal response
in the first layer is due to both the anisotropy of the molecular
potential and thermal excitations.

Using PIMC and a local superfluid density estimator, we find that
the first solvation layer surrounding a linear impurity is
approximately a two-dimensional superfluid.  Although this effect
has only a small contribution to the moment of inertia of this
system, it is clear that thermal excitations play a significant
role in the superfluid surrounding a molecule.  By varying the
temperature of the droplet, by adding $^3$He atoms, one could, in
principle, observe such temperature dependence in the moment of
inertia.

The authors would like to acknowledge K. Lehmann and B. Whaley for
useful discussions.  The computations were carried out at NCSA and the
IBM cluster at the Materials Computation Center, and was supported by
the NASA Microgravity Research Division, Fundamental Physics Program.
This work was also performed under the auspices of the U.S. Department
of Energy by University of California Lawrence Livermore National
Laboratory under contract No. W-7405-Eng-48.

$\dag$  Present Address:  Lawrence Livermore National Laboratory, 7000 East Avenue, L-415, Livermore, CA  94550.

\end{multicols}

\begin{references}

\bibitem{grebenev98} Grebenev, S., J.~P. Toennies and A.~F. Vilesov,
Science \textbf{279}, 2083 (1998).

\bibitem{callegari99} Callegari, C., A.~Conjusteau, I.~Reinhard,
K.~K. Lehmann, G.~Scoles and F.~Dalfovo, Phys. Rev. Lett. \textbf{83},
5058 (1999).

\bibitem{kwon99} Kwon, Y. and K.~B. Whaley,
Phys. Rev. Lett. \textbf{83}, 4108 (1999).

\bibitem{pollock87} Pollock, E.~L. and D.~M. Ceperley, Phys. Rev. B
\textbf{36}, 8343 (1987).

\bibitem{bishop81} Bishop, D.~J., J.~E. Berthold, J.~M. Parpia and
J.~D. Reppy, Phys. Rev. B \textbf{24}, 5047 (1981).

\bibitem{sindzingre89} Sindzingre, P., M.~L. Klein and D.~M. Ceperley,
Phys. Rev. Lett. \textbf{63}, 1601 (1989).

\bibitem{2dsd} Ceperley, D.~M., and Pollock, E.~L., Phys. Rev. B \textbf{39},
2084 (1989)

\bibitem{nyeki95} Ny{\'e}ki, J., R.~Ray, V.~Maidanov, M.~Siqueira,
B.~Cowan and J.~Saunders, J.  Low Temp. Phys. \textbf{101}, 279
(1995).

\bibitem{gordillo98} Gordillo, M.~C. and D.~M. Ceperley, Phys. Rev. B
\textbf{58}, 6447 (1998).

\bibitem{kim93} Kim, K. and W.~F. Saam, Phys. Rev. B \textbf{48},
13735 (1993).

\bibitem{nauta01} Nauta, K., private communication (2001).

\bibitem{nauta99} Nauta, K. and R.~E. Miller, Science \textbf{283}, 1895 (1999).


\bibitem{scoles92} Goyal, S., D.~L. Schutt and G.~Scoles, Phys. Rev. Lett. \textbf{69}, 933
  (1992).

\bibitem{pi99} Pi, M., R.~Mayol and M.~Barranco, Phys. Rev. Lett. \textbf{82}, 3093 (1999).

\bibitem{ceperley95} Ceperley, D.~M., Rev. Mod. Phys. \textbf{67}, 279 (1995).

\bibitem{lehmann98} Lehmann, K.~K., Mol. Phys. \textbf{97}, 645 (1998).

\bibitem{toennies98} Toennies, J.~P. and A.~F. Vilesov, Annu. Rev. Phys. Chem. \textbf{49}, 1
  (1998).

\bibitem{atkins96} Atkins, K.~M. and J.~M. Hutson, J. Chem. Phys. \textbf{105}, 440 (1996).

\bibitem{kwon00} Kwon, Y., P.~Huang, M.~V. Patel, D.~Blume and K.~B. Whaley, J. Chem. Phys.
  \textbf{113}, 6469 (2000).



\end{references}
\end{document}